\documentclass[fleqn,twoside]{article}
\usepackage{espcrc2}

\title{$K \rightarrow \pi\pi$ decay amplitudes from the lattice
\thanks{Presented by C.~Kim.  This work was supported by in part by
the U.~S.~Department of Energy.}}

\author{Changhoan Kim, Norman H.~Christ\address{Department of Physics, Columbia
University, New York, NY, 10027}}
\begin{document}


\begin{abstract}
In order to directly compute physical two-pion K-decay amplitudes using
lattice methods we must prepare a two-pion state with non-zero relative
momentum.  Building upon a proposal of Lellouch and L\"uscher, we describe
a finite-volume method to realize such a state as the lowest energy state
of two pions.
\end{abstract}

\maketitle

\section{Introduction}
\label{sec:Intro}

The techniques of lattice gauge theory offer the possibility of accurate,
non-perturbative calculation of the masses and matrix elements predicted
by low-energy QCD.  Of particular interest are the properties of the K-meson
system and the low-energy, weak matrix elements that are required to understand
Kaon decays, especially those which enter the $K \rightarrow \pi\pi$ amplitudes
which violate CP symmetry.  Presently well-developed methods permit the
calculation of $K$ to vacuum and $K$ to $\pi$ weak matrix elements.  When
combined with chiral perturbation theory and the accurate
chiral symmetry of the domain wall fermion formulation, lattice methods can
be used to evaluate all the matrix elements that contribute to the weak
$K \rightarrow \pi\pi$ amplitudes in the chiral limit\cite{Noaki:2001un,Blum:2001xb}.
In this paper we describe a method which may provide a practical
approach to the direct lattice calculation of the desired, on-shell,
$K \rightarrow \pi\pi$ amplitudes without the use of chiral perturbation
theory.

\section{Finite volume decay amplitudes}
We begin by reviewing the finite-volume approach developed by Lellouch and L\"uscher.
To compute the matrix element of an operator $O$ between two physical states
$|A\rangle$ and $|B\rangle$, one typically evaluates the Euclidean
Green's function:
\begin{displaymath}
G(t_1, t_2, t_3) = {\rm Tr}\Bigl\{ B(t_1) O(t_2) A(t_3) e^{-HT} \Bigr\}
\label{eq:euclid_green}
\end{displaymath}
in the limit $T \gg t_1 \gg t_2 \gg t_3$.
Here the operators $A$ and $B$ are chosen to carry quantum numbers which insure
that the desired states, $|A\rangle$ and $|B\rangle$ are each the lowest energy
states that can be created from the vacuum by $A$ and $B$, respectively.  As
demonstrated by Maiani and Testa\cite{Maiani:1990ca}, this method fails when one
of the desired states is a two-particle state above threshold.  In this case
the desired final state cannot be uniquely selected by its quantum numbers and
the Euclidean time limit will give an unwanted, threshold amplitude.

In a recent paper\cite{Lellouch:2000pv}, Lellouch and L\"uscher propose a method
to circumvent this difficulty by tuning the finite lattice volume so that the
first excited two-particle state allowed in that volume has the energy
of the initial decaying particle.  They work out an explicit formula which relates
the resulting finite-volume matrix element to the desired infinite volume decay
amplitude.  The same result is obtained by different methods by Lin {\it et al.},
offering further insight\cite{Lin:2001fi}.  As proposed, this method faces serious
practical difficulties given our present computational capabilities.  First, a
large spatial volume of $(6 \; {\rm Fermi})^3$ is required if the first-excited $\pi-\pi$
state is to have an energy matching the Kaon mass.  Second, the extraction of
the next-leading exponential in the large-time limit
is numerically difficult, especially if an accurate result is desired.  Finally,
while the matrix element of the $I=2$ final state can be obtained from the coefficient
of such a next-leading exponential, the more important $I=0$ state also receives
a vacuum contribution implying that the desired amplitude is the third term in
an expansion in increasingly small exponentials.

We propose to overcome these difficulties by imposing an anti-periodic boundary
condition on the pions, say in the $z$-direction, while using the usual periodic
boundary conditions in the other two directions.  With this choice the lowest
energy state of a single pion has momentum $p_z = {\pi \over L}$ instead of zero.
With such a choice of boundary conditions, the lowest energy state with
$I=2$ will be two pions, each with momentum $p_z \approx \pm{\pi \over L}$.  Thus, the
matrix element of such a state can be extracted from the leading exponential
in the Euclidean Green's function.  Since the vacuum state is unaffected by
these boundary conditions, our $I=0$ two-pion state will show the
next-leading large-time behavior and a vacuum subtraction will be necessary for
calculation of $\Delta I={1\over2}$ amplitudes.

The condition which insures that the two-pion state has the relative momentum
required for a physical $K \rightarrow \pi\pi$ decay, now requires
$L_z \approx 3$ Fermi, only one-half the lattice extent needed for the periodic
case.  The other two spatial dimensions are not constrained by these kinematics
and might be chosen to be similar $L_x=L_y=L_z$ giving a reduced box size of
$(3 \; {\rm Fermi})^3$.  Thus, the proposed method decreases the required spatial
size from $(6 \; {\rm Fermi})^3$ to $(3 \; {\rm Fermi})^3$ and promotes the amplitude
of interest to one of leading ($I=2$) or next-leading ($I=0$) large-time behavior.
Since only periodic and cubically symmetric periodic boundary conditions were
discussed in the original finite volume analysis\cite{Luscher:1991ux} on which
the Lellouch-L\"uscher paper\cite{Lellouch:2000pv} is based, we must recompute
L\"uscher's function $\phi(k)$ for this new, less symmetrical case.  This has
been done without difficulty following the method presented in the original
paper, Ref.~\cite{Luscher:1991ux}.

\section{G-parity boundary condition}

It is not immediately obvious how to realize this boundary condition for the
pion, because we have control only on the underlying quark fields. However,
the G-parity operation gives the solution since under G-parity:
$G | \pi > = -| \pi >$.  From this, we can see that anti-periodic boundary
conditions for the pion can be achieved by performing a G-parity operation at
the boundary---an operation that also can be defined at the quark level:
$u \rightarrow \bar{d}$ and $d \rightarrow -\bar{u}$.  (Recall that G is
defined as the product of charge-conjugation and an isospin rotation,
$G=C \exp{i\pi I_y}$.)  These unusual boundary were originally proposed
and implemented by Wiese\cite{Wiese:1992ku} and similar boundary conditions
have also been used in Ref.~\cite{Carmona:2000ds}.

Since we must also treat the Kaon, we must impose a charge conjugate boundary
condition on the strange quark for consistency.  This is easily done by
including a fictitious charmed quark, degenerate with the strange quark
and extending the usual G-parity by treating the $(c,s)$ quark pair as an
independent iso-doublet.  Our initial finite volume eigenstate can then be
constructed as a G-even mixture $(K^0+D^0)/\sqrt{2}$.
However, because our weak decay operator and our final states have definite
strangeness and charm, this mixing only introduces an extra
``finite-volume factor'' of $1/\sqrt{2}$ which can easily be removed.

\section{Lattice Action}

\subsection{Symmetry}
By imposing this G-parity boundary condition, we break chiral symmetry in a
manner similar to the chiral symmetry breaking of a mass term.  This was studied
for the continuum in Ref.~\cite{Wiese:1992ku}.  On the lattice, the boundary
condition is embedded in the action itself, adding a boundary-specific term.
For example, with G-parity boundary conditions we have a term in the action
density of the form:
\begin{equation}
\overline{u}(L_z-1) U^3(L_z-1) \gamma_3 C \overline{d}(0)^T,
\end{equation}
for Dirac conventions in which charge conjugation is represented by $q \rightarrow
C \overline{q}^T$.  Here $\overline{u}(L_z-1)$ is evaluated at the site $z=L_z-1$ and
$\overline{d}(0)$ at the site $z=0$ while $U^3(L_z-1)$ is the link variable joining
these two sites across the boundary.  Under chiral $SU(2)_L \otimes SU(2)_R$ such a
term transforms as a $({1 \over 2},{1 \over 2})$.  While such a boundary term may
appear to break translational invariance, the fact that the lattice Lagrangian is
invariant under the G-parity operation ensures us that the system has translational
symmetry, if appropriately defined.  (Note, when exploring unconventional boundary
conditions it is important to preserve the conservation of lattice momentum and hence
to require translational invariance.)

\subsection{Implementation}
The actual implementation of these boundary conditions represents a fairly minor
modification of the usual lattice Dirac operator.  An easy way of visualizing
these G-parity boundary conditions begins with doubling the lattice in the
$z$-direction, with the complex conjugates of gauge links in the original volume
copied onto the links translated by $L_z$ into this extended volume.  One then
easily defines a single-flavor Dirac operator on this $L_x\times L_y \times 2\cdot L_z$
lattice in the usual way, but imposing anti-periodic boundary conditions between
the $z=0$ and $z=2L_z-1$ boundaries.
If a quark field in the original volume is identified as a $u$ quark, then the
quark field in the doubled, $L_z \le z < 2L_z-1$ volume would be viewed as a
$\overline{d}$ quark.  Note, the computational load is twice that of a usual
calculation on a $L_x\times L_y \times L_z$ volume---our boundary conditions have
effectively required separate treatment for the $u$ and $d$ quarks.
Such an approach works for both Wilson and domain wall fermions.

\section{Conclusion}
\label{sec:Conclusion}
Since the G-parity boundary condition is imposed in only one direction, we lose
cubic symmetry.  In the cubically symmetric case the state of interest with orbital
angular momenta $l=0$ is mixed only with states having $l$ differing by multiples of 4.
However, without this cubic symmetry only parity invariance remains which permits
mixing with all even $l$.  Thus, in the practical application of the Lellouch and
L\"uscher method where the mixed, higher-$l$ states are treated as free and their
phase shifts ignored, we must neglect the $l=2$ phase shift $\delta_2(m_K)$.  This
reduces the accuracy of this approximation, introducing errors of order $(R/L)^4$
rather than $(R/L)^8$ as is the case for cubic symmetry\cite{Luscher:1991ux}.
Here $R$ is a typical strong interaction distance, {\it e.g.} $R=1/m_\rho$.

There is an additional complexity for the $I=0$ final state that should be discussed.
For simplicity, we consider a full QCD calculation which includes only dynamical $u$
and $d$ quarks.  For such a system, in addition to the three pions, there
will be a fourth $SU(2)$ singlet, quark-anti-quark state called here the
$\eta^\prime$.  This $\eta^\prime$ state is G-parity even in contrast to G-parity
odd triplet of pions and hence will obey even boundary conditions.  Thus an intended
two-pion, $I=0$ state may mix with an unwanted, threshold two-$\eta^\prime$
state which will have the same quantum numbers.  Fortunately, in the two-flavor
calculation just described, axial anomaly effects are expected make the non-Goldstone
$\eta^\prime$ heavier than the physical Kaon so this two-$\eta^\prime$ state will play
no role.  However, in a quenched calculation, this two-$\eta^\prime$ state will be
essentially degenerate with the desired $I=0$ two-pion state and obstruct the
calculation.  In fact, the hairpin diagrams entering the quenched two-$\eta^\prime$
amplitudes will give an extra power of $t$ or $t^2$ further obscuring the state of
interest.  Thus, this G-parity boundary condition can be used in the quenched
approximation only for the evaluation of $\Delta I =3/2$ amplitudes.

In summary, we propose imposing anti-periodic boundary conditions on the pion by
implementing a G-parity operation at boundary.  This reduces the required
simulation volume to $\approx (3 \; {\rm Fermi})^3$ and, more importantly, allows
the physical  $\langle \pi\pi(I)|K_{\rm weak}|K\rangle$ amplitude to be extracted
from the leading/next-leading large-time behavior of a Euclidean correlation
function for the $I=2/I=0$ final states.  We hope to try a calculation according
to this method in the coming year and thank our colleagues in the RBC collaboration
for useful suggestions and criticisms.


\end{document}